# Students' Perceptions and Attitudes towards the effectiveness of Prezi Uses in learning Islamic Subject

Azlina Mustaffa,
Master Student,
Institute of Education,,
International Islamic University Malaysia,
Gombak, Malaysia
azlina910@yahoo.com

Norazura Ezuana Mohd Najid,
Master Student
Institute of Education,
International Islamic University Malaysia,
Gombak, Malaysia
azuraezuana@yahoo.com

Siti Salwa Md. Sawari,
Master Student
Institute of Education,
International Islamic University Malaysia,
Gombak, Malaysia
salwa.sawari@gmail.com

*Abstract—* **Prezi is a Hungarian software company, producing a cloud-based presentation software and storytelling tool for presenting ideas on a virtual canvas (Prezi.com). Prezi is one of the teaching materials that help students in learning process. This study aims to explore the effectiveness of using Prezi in Islamic education subject among secondary schools under the topic of marriage in Islam: polygamy. Specifically its aims to identify students interest and second examine their attitude towards the uses of Prezi in learning Islamic educations. A total of 22 students participated in the survey, employing a 22-item questionnaire. The data was analyzed quantitatively using Statistical Package for the Social Sciences (SPSS). Result from this study revealed that student shows their interest in learning Islamic Educations when teachers uses Prezi. In additions students also show positive attitude towards uses of Prezi in Classroom. As conclusions, the uses of Prezi presentation is easy and its technique for developing a more creative and innovative approach in teaching strategies among Islamic educators.**

## I. INTRODUCTION

Today, we are facing students who are living in the world of digital natives. According to Marc Prensky (2001) and Platt (2011), screen teenagers are the screen natives and teachers are the screen immigrants. Today's students speak digitally; they are all "native speakers" of the digital language of computers (Xiao, 2013). If we educators do not take up the challenge to be compatible with our students' ability, we will be left behind. The circumstances will become more worst when the role of educator can as information transmitters will be no more significant since students can find everything they required without any guidance outside of their classrooms ( Jusoh, 2009).

If this happens, we as Muslim educators are responsible for it, based on Hadith of Prophet Muhammad (SAW):
"Every one of you is a caretaker, and every caretaker is responsible for what he is caretaker of" (Narrated by Imam Bukhari).

This study aims to explore the effectiveness of using Prezi in Islamic education subject among secondary schools under the topic of marriage in Islam: polygamy. There are different types of teaching aids involving computer usage or application such as PowerPoint presentation, macromedia flash, video, Facebook, website, Prezi and so on. Although different technologies such as moving film projectors, radio, instructional television, cassette players, and Video Cassette Recorders (VCRs) have been used in the classroom, they were only available around 1970s as personal computers found their way into schools (Sharp, 2006).

Generally, the materials that teachers use in teaching and learning process are to enhance students' performance in interest and perception towards the subject of Islamic education. This report will discuss about one of the teaching materials that help teachers in teaching and learning process (Forland & Kingston , 2004).

In order to make Islamic education class creative and innovative, teachers should know and be alert of the current technologies that are suitable and in line with students needs and interest. According to Keengwe (2008), technology cannot be assumed that once educational technology tools are available, teachers will have to integrate them into their daily classroom instruction. On the contrary, teachers need to go beyond the common task instead of just providing more machines in the classroom.

Because of the rapid technology, the researcher uses the latest application called Prezi. Prezi can be taught creatively in the classroom and through exciting activities in order to attract students' attention via zooming in the main ideas clearly and express clear explanation.

*A. Delimitation of The Study*

The study is focused on form five students in Sekolah Menengah Kebangsaan Seri Keramat, Gombak, Selangor Darul Ehsan. The researcher only taught one class (Form 5 Saidina Uthman) as a representative in order to examine the effectiveness of using Prezi in Islamic education subject especially in the topic polygamy.

## II. LITERATURE REVIEW

*A. Importance of Technology*

Nowadays in the urban life, the need of technology is more importance than ever. Studies have shown that with technology the students have shown their improvement in writing, math problem, social network and reading (Higgins, n.d). Furthermore, with technology the number of dropout rates has decrease and enchantment in their learning abilities .It also can improve student attendance and make the learning's more fun and exciting (Willingham, 2010). Moreover, via the technology the official procedures of school can be simplified and the school records such as the information about students, employs and teachers can be maintained effectively in a school database.

*B. Prezi*

The name of the product that researcher used was Prezi. Prezi the zooming editor tool was officially launched in April 2009 (Prezi.com). Prezi is a Hungarian software company, producing a cloud-based presentation software and storytelling tool for presenting ideas on a virtual canvas. The product employs a Zooming User Interface (ZUI), which allow users to zoom in and out of their presentation media, and allows user to display and navigate through information within a 2.5D or parallax 3D space on Z-axis. Prezi was officially established in 2009 by co-founders Adam Somlai –Fischer, Peter Halacsy and Peter Arvai. (refer www.prezi.com ).

This product was significant in new teaching and learning process in Islamic educaion. This is in line with Malaysia's aspiration to achieve students' abilities to increace their thinking critically and creatively.

The study also provide teachers with an alternative in teaching strategy in Islamic Education by using Prezi which consists of images, video clips and music. Transformation of education system today like *KSSM* emphasize on effective learning for students. According to A. H. Tamuri, 2007, he stated that most of Islamic teachers made minor modifications in approaches to the lessons, which relied on their own knowledge and creativity. In addition, problem-solving skills and ability to create new opportunities are suitable with Malaysia aspiration to develop excellent students.

The researcher included suitable images, Qur'anic verses with sound by famous readers, Sheikh Said AlGhamidi and two of the video clips that were appropriate to the topic. The first video was displayed during the set induction and last video during the conclusion to make students participate during question and answers session.

The process of making Prezi slide presentation is similar to creating a PowerPoint presentation. First, the researcher needs to go to www.prezi.com to download the programme to the computer. However, the differences of this method is in terms the focus of zooming and attracting the attention from the students.

First, when selecting the topic, the researcher used the topic 'polygamy' taken from form five subject of Islamic education. Content had been taken from textbook as it is the main reference of the students. Besides that, the researcher also searched for some information through internet and books to make sure students could understand the topic and the objective that the teacher wanted to achieve. All the information was included in Prezi in 15 slide presentation. This topic was taught for 30 minutes period. The researcher also chooses the template or concept of 'game' which shows metaphor or analogy when a husband starts to decide on polygamy as the solution in his life, he then needs to overcome all the obstacles in front him (see table 1).

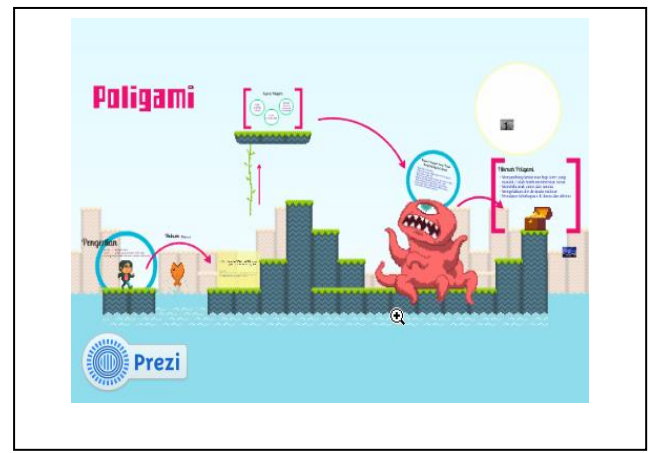

Table 1: General overview about the polygamy topic

Secondly, the researcher chose the video which is appropriate to the topic. It had been taken from Youtube, the famous channel where very useful and beneficial materials in the education field are shared or uploaded online. The researcher chosed the classical Malay famous movie, namely, *Madu Tiga* which is Tan Sri P. Ramlee as the actor. The video clip duration is approximately around three minutes to arouse students' attention. Next, the researcher chosed to write the definition of polygamy in second slides, the hukum of polygamy in third slide and also the quranic verses from surah An nisa, verses number three with the audio recitation by one of the famous qari Sheikh Said AlGhamidi from Masjidil Haram.

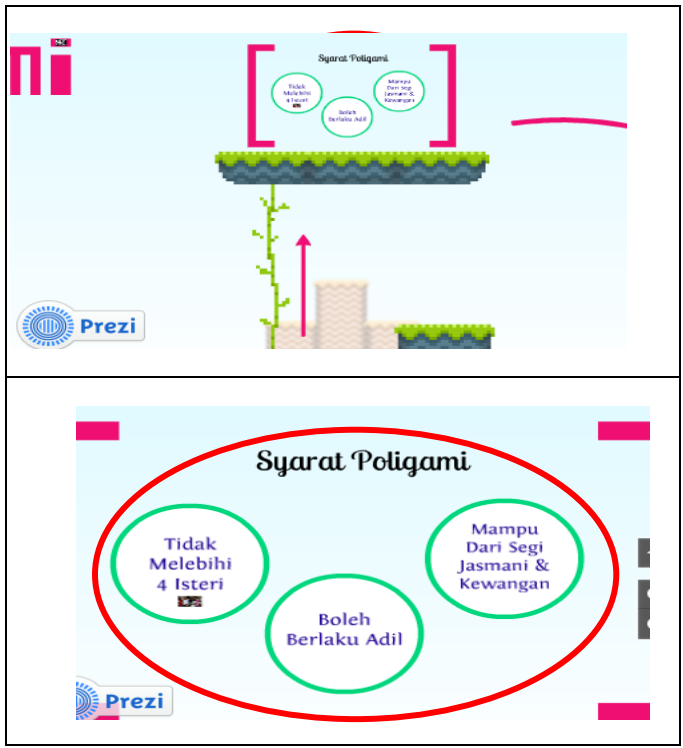

Table 2: Zoom in process

After that, the researcher selected the zoom in feature from slide 6 to 7 to give the students impact in focusing the important point about 'syarat poligami' and the explanation from the teacher. (See table 2).Furthermore, when the researcher started to discuss about the implication if the polygamy does not follow the Islamic rules, the researcher choosed the big obstacle for instance, monster images (Table 3) and direct zoom in to the main point.

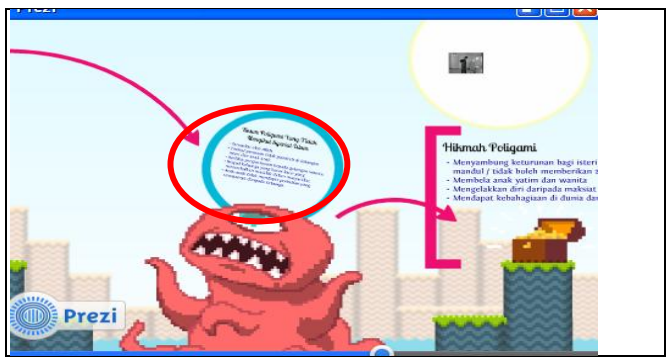

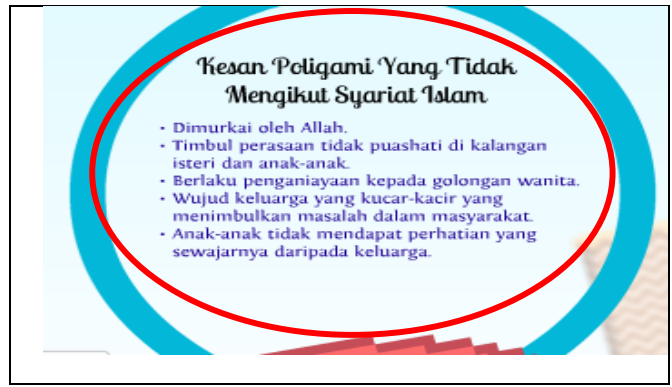

In addition, when the researcher wrote about the negative consequences, the slide after the obstacle shows about the advantages of polygamy in terms of individual and society.

In brief, the last presentation is about one of the video clips in the same film 'Madu Tiga' and the purposed of viewing the video clip is to attract the students' attention about some of the problems faced in the marriages and try to solved them by discussion with the teacher and students. As a conclusion of making the Prezi presentation, the researcher should make sure that the main objective will be achieved by considering the level of age and knowledge of the audience. It is important to ensure whether it is appropriate to the level of the teenagers or not.

*C. Students' Attitude]*

The uses of technology in learning cannot be denied nowadays.. Study done by Hong, Ridzuan and Kuek (2003) shows that most of students have a positive attitude towards technology as a learning tools. From their study, its successfully investigate that the uses of the technology can provide sufficient basic knowledge. In addition students also support the uses of the technology for learning. Initially the finding consistence with Wah (2006) when he discover that students considered the uses Instructional material give them benefit in improving their achievement in examination. This study involved Form Four students from selected secondary School. This result similarly being investigate by Cagiltary and Kaya (2011). The students which has been done on students (46) and alumni (21), 67 people (30 female, 37 male) from 12 different universities in Turkey discover that all of them records positive attitude towards the uses of technology in teaching and learning.

Moreover in It has been acknowledge that, students preferred technology as main support for their study (Usun, 2004). Usun run study on 156 undergraduate students from Department of Educational Sciences and Department of Computer and Educational Technologies in 2003 discovered that, students benefited from the uses of technology in learning.

## III. DATA COLLECTION

This section will explain how the data for this study are collected. The method is by using quantitative data via conducting a survey. The questionnaire distributed to 22 respondents from five Saidina Uthman. After the researcher presents the Prezi presentation to students, they were given the questionnaires which consist of three parts.

The first section is about students' interest towards using Prezi in Islamic education. For example, *I like to learn this subject by using software Prezi*. The second section is about students' perception towards teaching strategies like *Prezi usage in teaching process makes me maintain my concentration in learning Islamic studies* and the last part is about the tools in Prezi itself like the usage of fonts, music or video clips. For instance, *The usage of video clip, music, and pictures in Prezi are suitable to the topic*

## III. DATA ANALYSIS

After the questionnaires were answered by the students, the researcher analyses the data by using Statistical Package for the Social Sciences (SPSS). The data are presented in the format of frequencies and percentage. The researcher uses the SPSS programme to examine the result from the respondents. There are ten tables that shown ten items in the survey instrument.

|       |                | Frequency | Percent | Valid Percent | Cumulative Percent |
|-------|----------------|-----------|---------|---------------|--------------------|
| Valid | natural        | 1         | 4.5     | 4.5           | 4.5                |
|       | agree          | 7         | 31.8    | 31.8          | 36.4               |
|       | strongly agree | 14        | 63.6    | 63.6          | 100.0              |
|       | Total          | 22        | 100.0   | 100.0         |                    |

Table 1: Like to learn

For the first item which is about asking the students whether they *like to learn this subject by using software Prezi*. Table 1 as above shows that 63% of the respondents strongly agreed that Prezi could attract their interest, and that count was equivalent to 14 students. Another 7 students which was equivalent to 31% agreed that Prezi could increase their interest to learn.

|       |                | Frequency | Percent | Valid Percent | Cumulative Percent |
|-------|----------------|-----------|---------|---------------|--------------------|
| Valid | agree          | 5         | 22.7    | 22.7          | 22.7               |
|       | strongly agree | 17        | 77.3    | 77.3          | 100.0              |
|       | Total          | 22        | 100.0   | 100.0         |                    |

Table 2: Lessons in this subject are fun by using Prezi

Table 2 showed the percentage of the lessons in this subject are fun by using Prezi. 77% or equivalent to 17 students strongly agreed that the lesson was fun by using Prezi. Another 6 students or 20% of the respondents showed students strongly agreed this lesson was fun by using Prezi.

|       |                | Frequency | Percent | Valid Percent | Cumulative Percent |
|-------|----------------|-----------|---------|---------------|--------------------|
| Valid | agree          | 8         | 36.4    | 36.4          | 36.4               |
|       | strongly agree | 14        | 63.6    | 63.6          | 100.0              |
|       | Total          | 22        | 100.0   | 100.0         |                    |

Table 3: *This subject is one of my favourite school subjects by using Prezi*

Table 3 illustrated 63% of the students strongly agreed with the item, *this subject is one of my favourite school subjects by using Prezi* and that the percentage was equivalent to 14 students. The rest 36% of the students agreed and the count 8 students.

|       |                | Frequency | Percent | Valid Percent | Cumulative Percent |
|-------|----------------|-----------|---------|---------------|--------------------|
| Valid | natural        | 2         | 9.1     | 9.1           | 9.1                |
|       | agree          | 5         | 22.7    | 22.7          | 31.8               |
|       | strongly agree | 15        | 68.2    | 68.2          | 100.0              |
|       | Total          | 22        | 100.0   | 100.0         |                    |

Table 4: *Approach in this subject interest me in understanding of the content of subject via Prezi*

Table 4 showed the percentage of the approach in this subject interest in understanding of the content of subject via Prezi. Two of the students chosed to be neutral, meanwhile 5 students agreed and the rest 15 of the students strongly agreed that the use of Prezi makes them to understand the content well.

|  |  | Frequency | Percent | Valid Percent | Cumulative Percent |
|---|---|---|---|---|---|
| Valid | natural | 1 | 4.5 | 4.5 | 4.5 |
|  | agree | 4 | 18.2 | 18.2 | 22.7 |
|  | strongly agree | 17 | 77.3 | 77.3 | 100.0 |
|  | Total | 22 | 100.0 | 100.0 |  |

Table 5: *Prezi usage in teaching process makes me maintain my concentration in learning Islamic studies*

Table 5 showed the percentage of students that makes them to maintain their concentration in learning Islamic studies. It is showed that 17 of the students which is equivalent to 77% strongly agreed by using Prezi in teaching process make them maintain their concentration. The rest is 4 students, which is equivalent to 18% agreed that Prezi can make them maintain their concentration.

|  |  | Frequency | Percent | Valid Percent | Cumulative Percent |
|---|---|---|---|---|---|
| Valid | agree | 4 | 18.2 | 18.2 | 18.2 |
|  | strongly agree | 18 | 81.8 | 81.8 | 100.0 |
|  | Total | 22 | 100.0 | 100.0 |  |

Table 6: *The use of Prezi should be continued in teaching Islamic studies*

Table 6 shows the use of Prezi should be continued in teaching of Islamic studies. It shows that 4 students agreed and its equivalent to 18%. Another 18 students have strongly agreed that the use of Prezi should be continued in teaching of Islamic studies.

|  |  | Frequency | Percent | Valid Percent | Cumulative Percent |
|---|---|---|---|---|---|
| Valid | agree | 6 | 27.3 | 27.3 | 27.3 |
|  | strongly agree | 16 | 72.7 | 72.7 | 100.0 |
|  | Total | 22 | 100.0 | 100.0 |  |

Table 7: Prezi helps me to catch (understand) general overview about this topic

Table 7 shows the percentage of the catch (understanding) general overview about this topic. 16 students have strongly agreed that Prezi helps them to catch (understand) general overview about this topic which count for 72.7%. Another 6 students agreed to this item and it is equivalent to 27%.

|  |  | Frequency | Percent | Valid Percent | Cumulative Percent |
|---|---|---|---|---|---|
| Valid | natural | 1 | 4.5 | 4.5 | 4.5 |
|  | agree | 5 | 22.7 | 22.7 | 27.3 |
|  | strongly agree | 16 | 72.7 | 72.7 | 100.0 |
|  | Total | 22 | 100.0 | 100.0 |  |

Table 8: Usage of font

Table 8 shows the percentage regarding the text or writing is clearly could be read by students. 72% of the students strongly agreed that the text is clear in this Prezi, which was equivalent to 16 students. Meanwhile 5 students agreed that writing and text was readable which take 22% of the respondents. The rest, 1 student natural which counts for 4% the text was readable.

|  |  | Frequency | Percent | Valid Percent | Cumulative Percent |
|---|---|---|---|---|---|
| Valid | natural | 1 | 4.5 | 4.5 | 4.5 |
|  | agree | 14 | 63.6 | 63.6 | 68.2 |
|  | strongly agree | 7 | 31.8 | 31.8 | 100.0 |
|  | Total | 22 | 100.0 | 100.0 |  |

Table 9: I can answer the questions asked by teacher

Table 9 showed that the percentage of students could answer the questions asked by teacher. It showed that 14 of the students which was equivalent to 63% agreed could answer all the question given by the teacher. The rest was 7 of the students, which was equivalent to 31% strongly agreed that they could answer the question given by teacher.

|  |  | Frequency | Percent | Valid Percent | Cumulative Percent |
|---|---|---|---|---|---|
| Valid | agree | 1 | 4.5 | 4.5 | 4.5 |
|  | strongly agree | 21 | 95.5 | 95.5 | 100.0 |
|  | Total | 22 | 100.0 | 100.0 |  |

Table 10: The usage of video clip, music, and pictures in Prezi are suitable to the topic

Table 10 shows the usage of video clips, music, and pictures in Prezi was suitable to the topic or not. It shows 1 student agreed the suitability of using video clips, music, and pictures in Prezi and that is equivalent to 4.5%. Another 21 students strongly agreed about this video and counts for 95%.

| Reliability Statistics | |
|---|---|
| Cronbach's Alpha | N of Items |
| .883 | 10 |

Table 11: Reliability Test

From the SPSS, Table 11 shows the reliability statistics to determine the reliability; Cronbach's Alpha for the survey instrument was .883. The researcher could concluded that most of the form five students were satisfied with the use Prezi in Islamic education.

IV. CONCLUSION

Teaching and learning by using Prezi provides a new technique in terms of focusing and attracting students' attention in teaching and learning process. As a conclusion, the researcher may say that, by using a proper way of teaching and learning, we can achieve a certain improvement in the examination. Like what the researcher has done in this study, the students seemed to come from an average level of class. However, with the use of this method, it really provides students with positive attitude and interest towards learning Islamic education.

The method of using Prezi presentation is easy and its technique for developing a more creative and innovative approach in teaching strategies among Islamic educators. The method can be used not only in the teaching and learning in class and school but it can also be applied through seminar in which the presenters can explain clearly to the audience by using Prezi presentation.